\documentclass[12pt]{article}
\usepackage{graphicx}
\usepackage{epstopdf}
\usepackage{epsfig}
\usepackage{amsmath}
\usepackage{slashbox}
\usepackage{multirow}
\usepackage{subfigure}
\usepackage{rotating}
\usepackage{array}
\usepackage{amsfonts}
\def\lsim{\mathrel{\raise.3ex\hbox{$<$\kern-.75em\lower 1ex\hbox{$\sim$}}}}
\def\gsim{\mathrel{\raise.3ex\hbox{$>$\kern-.75em\lower 1ex\hbox{$\sim$}}}}

\def\be{\begin{equation}}
\def\ee{\end{equation}}
\def\bea{\begin{eqnarray*}}
\def\eea{\end{eqnarray*}}

\begin{document}
\title{One diagonal texture or cofactor zero \\
of the neutrino mass matrix}
\author{Jiajun Liao$^1$, D. Marfatia$^{2,3}$, and K. Whisnant$^1$\\
\\
\small\it $^1$Department of Physics and Astronomy, Iowa State University, Ames, IA 50011, USA\\
\small\it $^2$Department of Physics and Astronomy, University of Kansas, Lawrence, KS 66045, USA\\
\small\it $^3$Kavli Institute for Theoretical Physics, University of California, Santa Barbara, CA 93106, USA}
\date{}
\maketitle

\begin{abstract}

In view of the recent measurement of nonzero $\theta_{13}$, we carry
out a systematic study of a simple class of neutrino models that has
one diagonal texture or cofactor zero in the mass matrix. There are
seven free parameters in the model and five of them are already
measured by neutrino oscillation experiments; some cases for the
normal or inverted hierarchy are excluded and for the rest we obtain
the preferred values for the lightest neutrino mass and Dirac CP
phase. We find that there are
strong similarities between one diagonal texture zero models with
one mass hierarchy and one diagonal cofactor zero models with the
opposite mass hierarchy. We also make predictions for neutrinoless
double beta decay for these models. For the one
cofactor zero models, we present a simple realization based on a new
$U(1)$ gauge symmetry.
\end{abstract}

\newpage
\section{Introduction}

After the recent observation of nonzero $\theta_{13}$ by the Daya
Bay~\cite{An:2012eh}, RENO~\cite{Ahn:2012nd}, and Double
Chooz~\cite{DC} experiments, five parameters in the neutrino sector
have been measured by neutrino oscillation experiments. In general,
there are nine parameters in the light neutrinos mass matrix. The
remaining four unknown parameters may be taken as the lightest mass,
the Dirac CP phase and two Majorana phases. The Dirac phase will be
measured in future long baseline neutrino experiments, and the
lightest mass can be determined from beta decay and cosmological
experiments. If neutrinoless double beta decay ($0\nu\beta\beta$) is
detected, a combination of the two Majorana phases can also be probed.
If there is some structure in the neutrino mass matrix, the four unknown
parameters will be related to each other. In this paper, we study the
phenomenological consequence of imposing one texture zero or one
cofactor zero in the light neutrino mass matrix; for previous work see
Refs.~\cite{Frampton:2002yf, Lavoura:2004tu, Liao:2013kix,
  Araki:2012ip}. Since one texture or cofactor zero sets two
conditions on the parameter space, only seven free parameters in the
light neutrino matrix remain. Here we derive analytic formulas that
relate the seven free parameters and determine the constraints on
these models. By using recent data measured by neutrino oscillation
experiments, we exclude some cases for the normal or inverted mass
hierarchy, and for the rest we can obtain the allowed regions for the
lightest mass and Dirac CP phase, which can be probed in the next
generation of neutrino experiments.

In Sec. 2, we discuss the general properties of texture or cofactor
zeros in the light neutrino mass matrix. In Sec. 3, we use current
experimental data to study the allowed parameter regions for one
diagonal texture zero in the mass matrix. In Sec. 4, we study the
allowed parameter regions for one diagonal cofactor zero in the mass
matrix. In Sec. 5 we discuss the similarities between one texture zero
models with one mass hierarchy and one cofactor zero models with the
opposite mass hierarchy. We present a simple realization based on a
new $U(1)$ gauge symmetry for the one cofactor zero models in Sec. 6
and summarize our results in Sec. 7.

\section{General properties of texture or cofactor zeros of the
neutrino mass matrix}

The light neutrino mass matrix can be written as 
\begin{equation}
M=V^*\text{diag}(m_1, m_2, m_3)V^\dagger\,,
\label{eq:Mnu}
\end{equation}
where $V=U\text{diag}(1, e^{i\phi_2/2}, e^{i\phi_3/2})$, and 
\begin{align}
U=\begin{bmatrix}
   c_{13}c_{12} & c_{13}s_{12} & s_{13}e^{-i\delta} \\
   -s_{12}c_{23}-c_{12}s_{23}s_{13}e^{i\delta} & c_{12}c_{23}-s_{12}s_{23}s_{13}e^{i\delta} & s_{23}c_{13} \\
   s_{12}s_{23}-c_{12}c_{23}s_{13}e^{i\delta} & -c_{12}s_{23}-s_{12}c_{23}s_{13}e^{i\delta} & c_{23}c_{13}
   \end{bmatrix}.
\label{eq:U}
\end{align}
The results of a recent global three-neutrino fit~\cite{Fogli:2012ua} are
shown in Table~1. The masses of light neutrinos can be obtained from
the canonical seesaw mechanism~\cite{seesaw}, in which
the mass matrix of the light neutrinos can be written as
\begin{align}
M=Y_\nu^T M_R^{-1}Y_\nu v^2\,,
\label{eq:seesaw}
\end{align}
where $v\approx 174$ GeV is the Higgs vacuum expectation value (VEV),
$Y_\nu$ is the $3\times 3$ Yukawa coupling matrix and $M_R$ is the
$3\times 3$ heavy right-handed neutrino mass matrix. Here we assume
all three light neutrinos are massive, so that the mass matrix of
the light neutrinos is invertible (and therefore $Y_\nu$ must be
invertible), and we can write Eq.~(\ref{eq:seesaw}) as
\begin{align}
M_R = Y_\nu M^{-1} Y_\nu^T v^2\,.
\end{align}
Since $(M^{-1})_{\alpha\beta}=\frac{1}{\det M}C_{\beta\alpha}$, where
$C_{\beta\alpha}$ is the $(\beta,\alpha)$ cofactor of $M$, and both
the light and heavy neutrino mass matrices are symmetric, any cofactor
zeros in the mass matrix are equivalent to texture zeros in the
inverse of the mass matrix.  Consequently, Eq.~(4) implies that if the
Yukawa coupling matrix is diagonal, then a cofactor zero in $M$
implies a texture zero in $M_R$~\cite{Ma:2005py}. Similarly, a texture
zero in $M$ implies a cofactor zero in $M_R$ when the Yukawa coupling
matrix is diagonal.

\begin{table}
\caption{Best-fit values and $2\sigma$ ranges of the oscillation
parameters~\cite{Fogli:2012ua}, with $\delta m^2 \equiv m_2^2-m_1^2$
and $\Delta m^2 \equiv m_3^2-(m_1^2+m_2^2)/2$.
\label{tab:data}}
\begin{center}
\begin{tabular}{|l|*{5}{c|}}\hline
\makebox[4em]{Hierarchy}
&\makebox[2em]{$\theta_{12}(^\circ)$}&\makebox[2em]{$\theta_{13}(^\circ)$}&\makebox[2em]{$\theta_{23}(^\circ)$}
&\makebox[6em]{$\delta m^2(10^{-5}\text{eV}^2)$}&\makebox[7em]{$|\Delta m^2|(10^{-3}\text{eV}^2)$}\\\hline
Normal &$33.6^{+2.1}_{-2.0}$&$8.9^{+0.9}_{-0.9}$&$38.4^{+3.6}_{-2.3}$&$7.54^{+0.46}_{-0.39}$&$2.43^{+0.12}_{-0.16}$\\\hline
Inverted &$33.6^{+2.1}_{-2.0}$&$9.0^{+0.8}_{-1.0}$&$38.8^{+5.3}_{-2.3}\oplus 47.5-53.2$&$7.54^{+0.46}_{-0.39}$&$2.42^{+0.11}_{-0.16}$\\\hline
\end{tabular}
\end{center}
\end{table}

An interesting feature of the structure of a texture or cofactor zero
is that it is stable against radiative corrections. The one-loop
renormalization group equation (RGE) describing the evolution of the
light neutrino masses from the lightest right-handed neutrino mass
scale $M_1$ to the electroweak scale $M_Z$ is~\cite{Antusch:2005gp}
\begin{equation}
16\pi^2\frac{dM}{dt}=\alpha M + C[(Y_lY_l^\dagger)M+M(Y_lY_l^\dagger)^T],
\label{eq:RGE}
\end{equation}
where $t = \ln(\mu/M_1)$, $\mu$ is the renormalization scale and
$Y_l=\text{diag}(y_e, y_\mu, y_\tau)$ is the charged lepton Yukawa
coupling matrix. In the Standard Model (SM), $C=-\frac{3}{2}$ and
$\alpha\approx-3g_2^2+6y_t^2+\lambda$, and in the minimal
supersymmetric standard model, $C=1$ and
$\alpha\approx-\frac{6}{5}g_1^2-6g_2^2+6y_t^2$, where $g_1,g_2$ are
the gauge couplings, $y_t$ is the top quark Yukawa coupling, and
$\lambda$ is the Higgs self-coupling. The solution to
Eq.~(\ref{eq:RGE}) can be written as~\cite{solution}
\begin{equation}
M(M_Z)=I_\alpha\begin{bmatrix}
   I_e & 0 & 0 \\
   0 & I_\mu & 0  \\
   0 & 0 & I_\tau
   \end{bmatrix}M(M_1)\begin{bmatrix}
      I_e & 0 & 0 \\
      0 & I_\mu & 0  \\
      0 & 0 & I_\tau
      \end{bmatrix},
\label{eq:RGEsolution}
\end{equation}
where 
\begin{equation}
I_\alpha=\exp\left[-\frac{1}{16\pi^2}\int_0^{\ln(M_1/M_Z)}\alpha(t)dt\right],
\end{equation}
and 
\begin{equation}
I_l=\exp\left[-\frac{C}{16\pi^2}\int_0^{\ln(M_1/M_Z)}y_l^2(t)dt\right],
\end{equation}
for $l=e,\mu,\tau$. Since multiplying diagonal matrices does not
affect a texture or cofactor zero, from Eq.~(\ref{eq:RGEsolution}), we
see that texture or cofactor zero models are stable
against the RGE running from $M_1$ to $M_Z$.

\section{One texture zero in the neutrino mass matrix}

We discuss the properties of one texture zero in the
diagonal entries of the mass matrix; the results for the off-diagonal
cases can be found in Ref.~\cite{Liao:2013kix}, which were obtained in
models with four texture zeros in the Yukawa coupling matrix.

Our analysis proceeds as follows. For one texture zero cases, there
are 7 independent parameters in the light neutrino mass matrix, which
we take to be $\theta_{12}$, $\theta_{23}$, $\theta_{13}$, $\delta
m^2$, $\Delta m^2$, the Dirac CP phase $\delta$, and either $m_1$ (for
the normal hierarchy, NH, $m_1<m_2<m_3$) or $m_3$ (for the inverted
hierarchy, IH, $m_3<m_1<m_2$). For each case we find the allowed
regions in the $m_1$- ($m_3$-) $\delta$ plane given the best-fit
values of $\theta_{12}$, $\theta_{23}$, $\theta_{13}$, $\delta m^2$
and $\Delta m^2$, and also the $2\sigma$ allowed regions using the
experimental uncertainties in the measured parameters. We also find
iso-$|M_{ee}|$ contours relevant for neutrinoless double beta decay
for the best-fit values.

\subsection{$M_{ee}=0$}

The condition $M_{ee}=0$ can be written as
\begin{equation}
m_1 = - \frac{m_3e^{i\phi_3}U_{e3}^2+m_2e^{i\phi_2}U_{e2}^2}{U_{e1}^2},
\label{eq:condition_M11}
\end{equation}
and is the same for either mass hierarchy. Taking the absolute square gives
\begin{equation}
m_1^2|U_{e1}|^4-m_2^2|U_{e2}|^4-m_3^2|U_{e3}|^4
= 2{\rm Re}(m_3 e^{-i\phi_3} U_{e3}^{*2} m_2 e^{i\phi_2} U_{e2}^2),
\end{equation}
or, defining $\phi = \phi_3-\phi_2$, 
\begin{equation}
m_1^2|U_{e1}|^4-m_2^2|U_{e2}|^4-m_3^2|U_{e3}|^4= 2 m_2 m_3 c_{13}^2 s_{13}^2 s_{12}^2\cos(-\phi+2\delta).
\label{eq:M11c}
\end{equation}
Expanding the cosines yields the form
\begin{align}
C = A\cos\phi + B\sin\phi,
\label{eq:condition}
\end{align}
with A, B and C as listed in Table~\ref{tab:ABC}. Hence the only
condition that must be satisfied when $M_{ee}=0$ is $C^2 \leq A^2 +
B^2$. Since $C^2$ and $A^2 + B^2$ do not depend on $\delta$, it will
only yield a constraint on $m_1$ for the normal hierarchy or $m_3$ for
the inverted hierarchy. It can be easily seen that $C^2 \leq A^2 + B^2$
cannot be satisfied for the inverted hierarchy, which means that
$M_{ee}=0$ is not possible for the inverted hierarchy. For the normal
hierarchy and best-fit oscillation parameters, the allowed range for
$m_1$ is $0.0022 \text{ eV}\leq m_1 \leq 0.0066 \text{ eV}$, while the
allowed range at 2$\sigma$ is $0.0014 \text{ eV}\leq m_1\leq 0.0085
\text{ eV}$.

\begin{sidewaystable}
\centering
$C=A\cos\phi + B\sin\phi$\\
\begin{tabular}{|c|c|c|c|}\hline
Class & A & B & C \\
\hline
$M_{ee}=0$&$2 m_2 m_3 c_{13}^2 s_{13}^2 s_{12}^2 \cos(2\delta)$&$2 m_2 m_3 c_{13}^2 s_{13}^2 s_{12}^2 \sin(2\delta)$&$m_1^2|U_{e1}|^4-m_2^2|U_{e2}|^4-m_3^2|U_{e3}|^4$\\
\hline
$M_{\mu\mu}=0$&$\begin{array}{l}
2 m_2 m_3 s_{23}^2 c_{13}^2 \times\\
\left[c_{12}^2c_{23}^2+s_{12}^2s_{23}^2s_{13}^2\cos (2\delta)\right.\\
\left.-2c_{12}s_{12}c_{23}s_{23}s_{13}\cos \delta\right]
\end{array}$&$\begin{array}{l} 
2 m_2 m_3 s_{23}^2 c_{13}^2 \times\\
\left[s_{12}^2s_{23}^2s_{13}^2\sin (2\delta)\right.\\ 
\left.-2c_{12}s_{12}c_{23}s_{23}s_{13}\sin\delta\right]
\end{array}$&$ m_1^2|U_{\mu 1}|^4-m_2^2|U_{\mu 2}|^4-m_3^2|U_{\mu 3}|^4$\\
\hline
$M_{\tau\tau}=0$&$\begin{array}{l}
2 m_2 m_3 c_{23}^2 c_{13}^2  \times\\
\left[c_{12}^2s_{23}^2+s_{12}^2c_{23}^2s_{13}^2\cos (2\delta)\right.\\
\left.+2c_{12}s_{12}c_{23}s_{23}s_{13}\cos \delta\right]
\end{array}$&$\begin{array}{l} 
2 m_2 m_3 c_{23}^2 c_{13}^2 \times\\
\left[s_{12}^2c_{23}^2s_{13}^2\sin (2\delta)\right.\\ 
\left.+2c_{12}s_{12}c_{23}s_{23}s_{13}\sin\delta\right]
\end{array}$&$ m_1^2|U_{\tau 1}|^4-m_2^2|U_{\tau 2}|^4-m_3^2|U_{\tau 3}|^4$\\
\hline
$C_{ee}=0$&$2 m_2^{-1} m_3^{-1} c_{13}^2 s_{13}^2 s_{12}^2 \cos(2\delta)$&$2 m_2^{-1} m_3^{-1} c_{13}^2 s_{13}^2 s_{12}^2 \sin(2\delta)$&$m_1^{-2}|U_{e1}|^4-m_2^{-2}|U_{e2}|^4-m_3^{-2}|U_{e3}|^4$\\
\hline
$C_{\mu\mu}=0$&$\begin{array}{l}
2 m_2^{-1} m_3^{-1} s_{23}^2 c_{13}^2 \times\\
\left[c_{12}^2c_{23}^2+s_{12}^2s_{23}^2s_{13}^2\cos (2\delta)\right.\\
\left.-2c_{12}s_{12}c_{23}s_{23}s_{13}\cos \delta\right]
\end{array}$&$\begin{array}{l} 
2 m_2^{-1} m_3^{-1} s_{23}^2 c_{13}^2 \times\\
\left[s_{12}^2s_{23}^2s_{13}^2\sin (2\delta)\right.\\ 
\left.-2c_{12}s_{12}c_{23}s_{23}s_{13}\sin\delta\right]
\end{array}$&$ m_1^{-2}|U_{\mu 1}|^4-m_2^{-2}|U_{\mu 2}|^4-m_3^{-2}|U_{\mu 3}|^4$\\
\hline
$C_{\tau\tau}=0$&$\begin{array}{l}
2 m_2^{-1} m_3^{-1} c_{23}^2 c_{13}^2  \times\\
\left[c_{12}^2s_{23}^2+s_{12}^2c_{23}^2s_{13}^2\cos (2\delta)\right.\\
\left.+2c_{12}s_{12}c_{23}s_{23}s_{13}\cos \delta\right]
\end{array}$&$\begin{array}{l} 
2 m_2^{-1} m_3^{-1} c_{23}^2 c_{13}^2 \times\\
\left[s_{12}^2c_{23}^2s_{13}^2\sin (2\delta)\right.\\ 
\left.+2c_{12}s_{12}c_{23}s_{23}s_{13}\sin\delta\right]
\end{array}$&$ m_1^{-2}|U_{\tau 1}|^4-m_2^{-2}|U_{\tau 2}|^4-m_3^{-2}|U_{\tau 3}|^4$\\
\hline
\end{tabular}
\caption{The coefficients A, B and C for each class.
\label{tab:ABC}}
\end{sidewaystable}

\subsection{$M_{\mu\mu}=0$}

\begin{figure}
\centering
\includegraphics[width=6.0in]{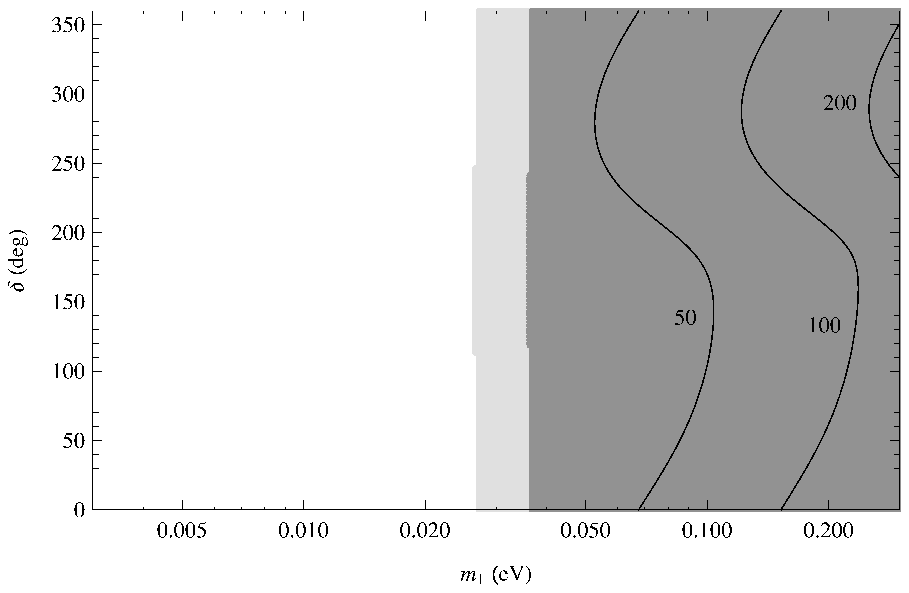}
\caption{The allowed regions in the $(m_1,\delta)$ plane for $M_{\mu\mu}=0$
and the normal hierarchy. The dark shaded regions correspond to the best-fit
oscillation parameters, while the light shaded regions
are allowed at $2\sigma$. The solid lines are iso-$|M_{ee}|$ contours
(in meV).}
\label{fg:M22-NO}
\end{figure}

\begin{figure}
\centering
\includegraphics[width=6.0in]{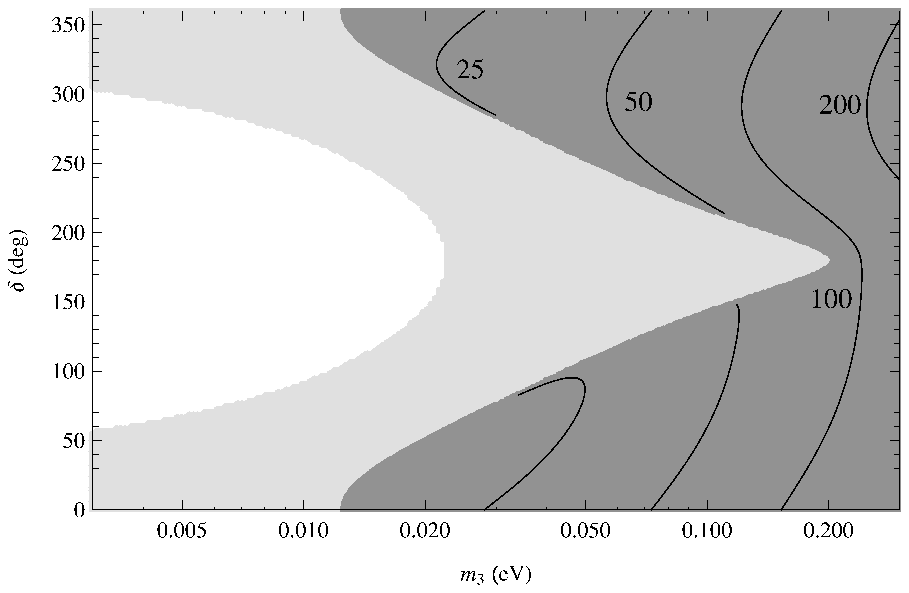}
\caption{Same as Fig.~\ref{fg:M22-NO}, except for $M_{\mu\mu}=0$ and the
inverted hierarchy.}
\label{fg:M22-IO}
\end{figure}

From $M_{\mu\mu} = 0$, we get
\begin{equation}
m_1 = - \frac{m_3e^{i\phi_3}U_{\mu 3}^2+m_2e^{i\phi_2}U_{\mu 2}^2}{U_{\mu 1}^2}\,,
\label{eq:condition_M22}
\end{equation}
which is independent of hierarchy. As before this may be put in the
form of Eq.~(\ref{eq:condition}), with A, B and C given in
Table~\ref{tab:ABC}. From Eq.~(\ref{eq:condition}), we can find the
solution
\begin{equation}
\phi=2\arctan\frac{B \pm \sqrt{A^2+B^2-C^2}}{A+C},
\label{eq:phi}
\end{equation}
and we can write Eq.~(\ref{eq:condition_M22}) as
\begin{equation}
m_1 =e^{i\phi_2}\frac{-m_3 e^{i\phi}U_{\mu 3}^2-m_2 U_{\mu 2}^2} {U_{\mu 1}^2}.
\end{equation}
Since $m_1$ is a non-negative real number in the standard parametrization, we get
\begin{equation}
\phi_2=-\text{arg}[\frac{-m_3 e^{i\phi}U_{\mu 3}^2- m_2 U_{\mu 2}^2} {U_{\mu 1}^2}],
\end{equation}
and
\begin{equation}
\phi_3=\phi_2+\phi.
\end{equation}
It is then possible to calculate the magnitude of the $\nu_e-\nu_e$
element of the neutrino mass matrix
\begin{equation}
|M_{ee}|=|m_1c_{12}^2c_{13}^2+m_2 e^{-i\phi_2}s_{12}^2c_{13}^2+m_3 e^{-i\phi_3}s_{13}^2e^{2i\delta}|,
\end{equation}
which determines the rate for neutrinoless double-beta decay, a signal
of lepton number violation. The allowed regions of the Dirac CP phase
$\delta$ and the lightest mass $m_1$ ($m_3$) are defined by 
the condition $C^2 \leq A^2 + B^2$. We scan over
$\delta$ and $m_1$ ($m_3$) to find the allowed regions; see
Fig.~\ref{fg:M22-NO} for the normal hierarchy and Fig.~\ref{fg:M22-IO} for
the inverted hierarchy, where regions corresponding to the best-fit
parameters and those allowed at $2\sigma$ are shown. The lightest mass
for the normal hierarchy is always larger than 0.027~eV at $2\sigma$,
while for the inverted hierarchy, it is strongly dependent on $\delta$.
We also plot iso-$|M_{ee}|$ contours using the
best-fit oscillation parameters. Here only the contours for the plus
sign of $\phi$ in Eq.~(\ref{eq:phi}) are shown because changing
$\delta$ to $360^\circ -\delta$ yields the same contours for the minus
solution.

\subsection{$M_{\tau\tau}=0$}

\begin{figure}
\centering
\includegraphics[width=6.0in]{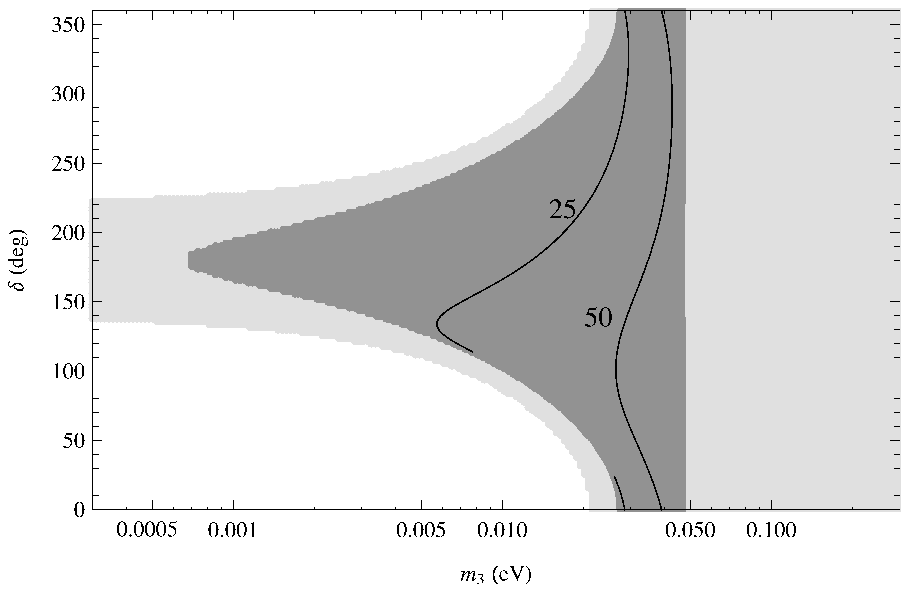}
\caption{Same as Fig.~\ref{fg:M22-NO}, except for $M_{\tau\tau}=0$ and the
inverted hierarchy.}
\label{fg:M33-IO}
\end{figure}

From $M_{\tau\tau} = 0$, we get
\begin{equation}
m_1 = - \frac{m_3e^{i\phi_3}U_{\tau 3}^2+m_2e^{i\phi_2}U_{\tau 2}^2}{ U_{\tau 1}^2}\,,
\label{eq:condition_M33}
\end{equation}
which is independent of hierarchy. This may be put in the form of
Eq.~(\ref{eq:condition}), with A, B and C as in Table~\ref{tab:ABC}.
We find that the normal hierarchy is excluded at $2\sigma$. The
allowed regions for the inverted hierarchy are shown in
Fig.~\ref{fg:M33-IO}, along with iso-$|M_{ee}|$ contours. Note
that for the best-fit oscillation parameters, the lightest mass $m_3$
has an upper bound of 0.047~eV, but there is no upper bound at
$2\sigma$.

\section{One cofactor zero of the neutrino mass matrix}

We now discuss the properties of one cofactor zero in the diagonal
entries of the mass matrix; the results for the off-diagonal cases can
be found in Ref.~\cite{Liao:2013kix}, which were obtained in models
with four texture zeros in the Yukawa coupling matrix.  Our analysis
follows the same procedure as for the texture zeros in the previous
section.

\subsection{$C_{ee}=0$}

If $C_{ee}=0$, then $(M^{-1})_{ee}=0$. Since
$M^{-1}=V\text{diag}(m_1^{-1}, m_2^{-1}, m_3^{-1})V^T$, we can write
the condition as
\begin{equation}
m_1^{-1} = - \frac{m_3^{-1}e^{i\phi_3}U_{e3}^2+m_2^{-1}e^{i\phi_2}U_{e2}^2}{U_{e1}^2}\,,
\label{eq:condition_C11}
\end{equation}
which is the same for either mass hierarchy. Taking the absolute
square, we write this in the form of Eq.~(\ref{eq:condition}), with A,
B and C as in Table~\ref{tab:ABC}. Since $C^2$ and $A^2 + B^2$ do not
depend on $\delta$, it will only yield a constraint on $m_1$ ($m_3$)
for the normal (inverted) hierarchy. We find that the normal hierarchy
is excluded at 2$\sigma$. For the inverted hierarchy and best-fit
oscillation parameters, the allowed range for $m_3$ is $0.0013 \text{
  eV}\leq m_3 \leq 0.0031 \text{ eV}$, while the allowed range at
2$\sigma$ is $0.0010 \text{ eV}\leq m_3\leq 0.0042 \text{ eV}$.

\subsection{$C_{\mu\mu}=0$}

\begin{figure}
\centering
\includegraphics[width=6.0in]{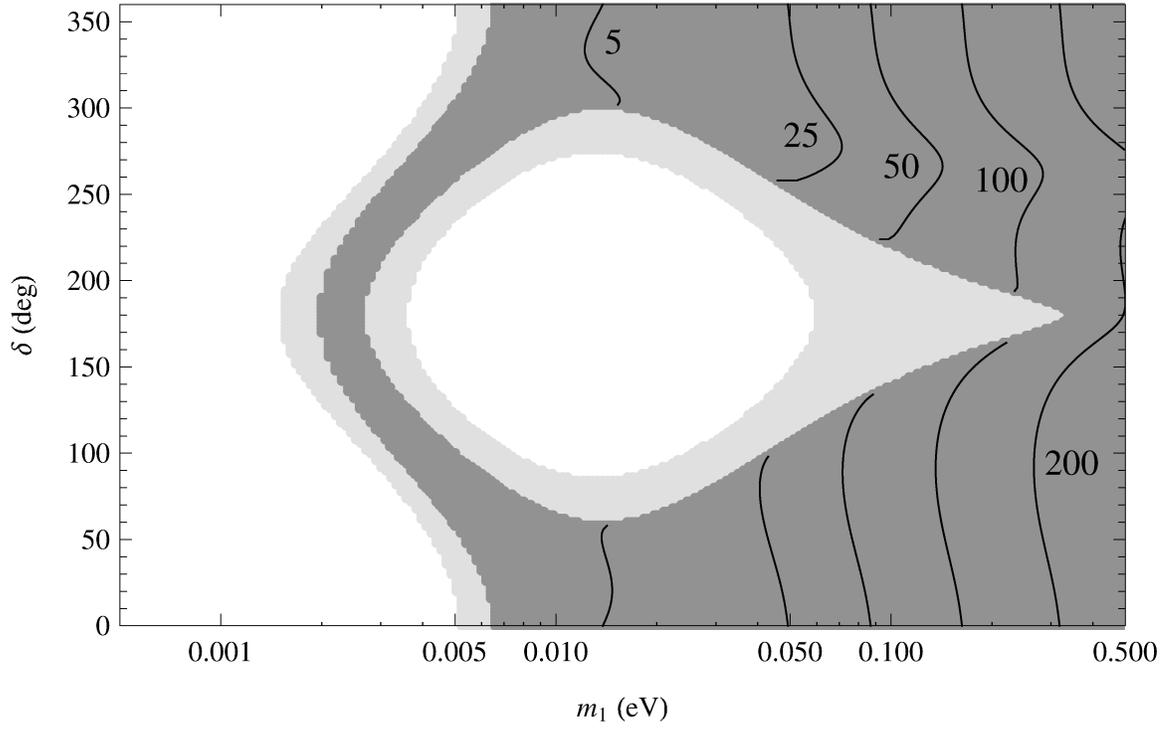}
\caption{Same as Fig.~\ref{fg:M22-NO}, except for $C_{\mu\mu}=0$ and the
normal hierarchy.}
\label{fg:C22-NO}
\end{figure}

\begin{figure}
\centering
\includegraphics[width=6.0in]{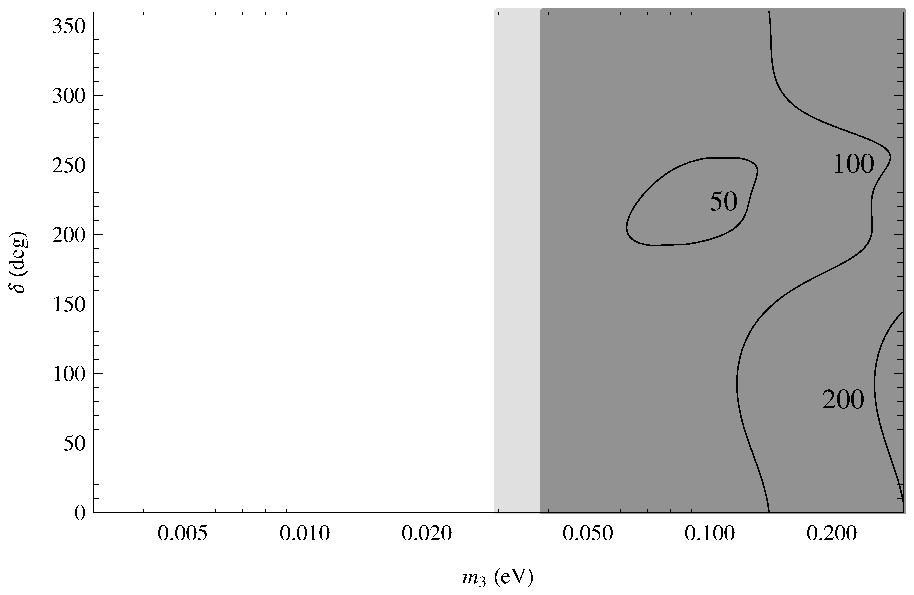}
\caption{Same as Fig.~\ref{fg:M22-NO}, except for $C_{\mu\mu}=0$ and the
inverted hierarchy.}
\label{fg:C22-IO}
\end{figure}

From $C_{\mu\mu} = 0$, which is equivalent to $(M^{-1})_{\mu\mu}=0$, we get
\begin{equation}
m_1^{-1} = -\frac{m_3^{-1}e^{i\phi_3}U_{\mu 3}^2+m_2^{-1}e^{i\phi_2}U_{\mu 2}^2}{U_{\mu 1}^2}\,,
\label{eq:condition_C22}
\end{equation}
which is the same for either mass hierarchy, and may be put in the
form of Eq.~(\ref{eq:condition}), with A, B and C as in
Table~\ref{tab:ABC}.  The allowed regions for the normal hierarchy are
shown in Fig.~\ref{fg:C22-NO} and the allowed regions for the inverted
hierarchy are shown in Fig.~\ref{fg:C22-IO}.

\subsection{$C_{\tau\tau}=0$}

\begin{figure}
\centering
\includegraphics[width=6.0in]{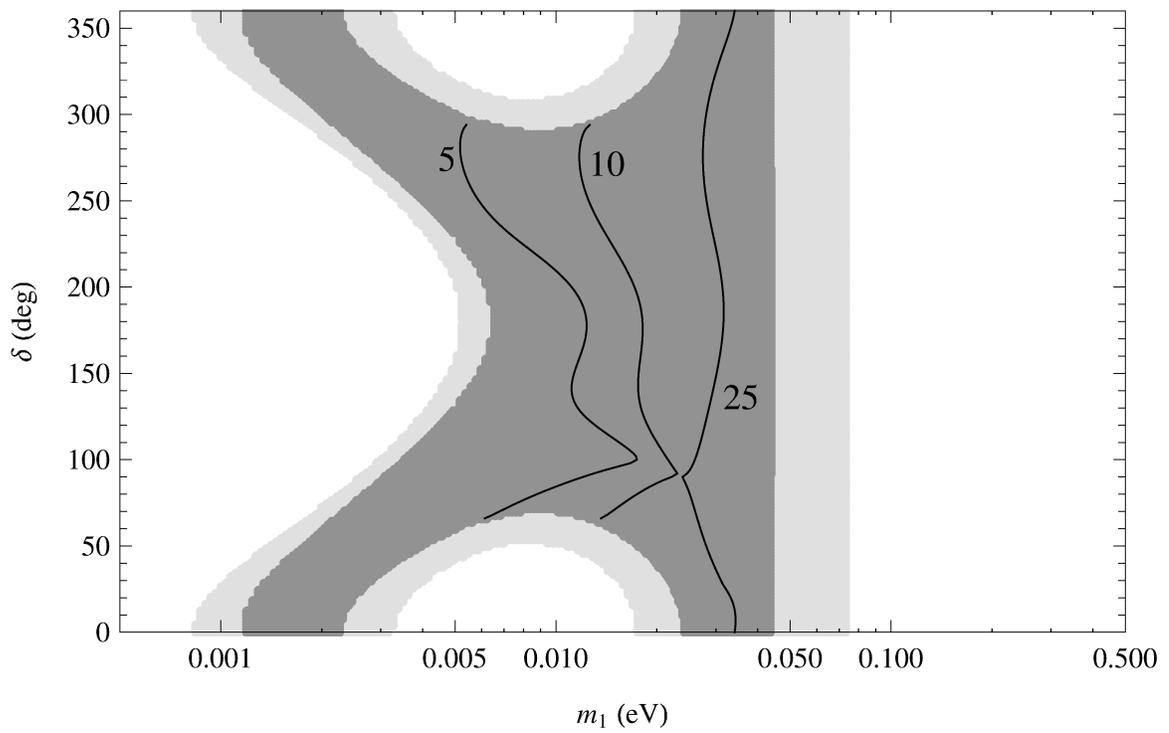}
\caption{Same as Fig.~\ref{fg:M22-NO}, except for $C_{\tau\tau}=0$ and the
normal hierarchy.}
\label{fg:C33-NO}
\end{figure}

From $C_{\tau\tau} = 0$, which is equivalent to $(M^{-1})_{\tau\tau}=0$, we get
\begin{equation}
m_1^{-1} = -\frac{m_3^{-1}e^{i\phi_3}U_{\tau 3}^2+m_2^{-1}e^{i\phi_2}U_{\tau 2}^2}{U_{\tau 1}^2}.
\label{eq:condition_C33}
\end{equation}
This condition is the same for either mass hierarchy, and may be put
in the form of Eq.~(\ref{eq:condition}), with A, B and C as in
Table~\ref{tab:ABC}. We find that for the inverted hierarchy, this case is
not allowed for the best-fit parameters, but is allowed at
2$\sigma$, with a lower bound on $m_3$ of 0.033~eV. The allowed regions
for the normal hierarchy are shown in
Fig.~\ref{fg:C33-NO}, along with iso-$|M_{ee}|$ contours. We see that
the lightest mass $m_1$ has an upper bound of 0.044~eV for the
best-fit oscillation parameters, and 0.071~eV at $2\sigma$.

\section{Similarity of texture-zero and cofactor-zero models}

The allowed regions for Class $C_{\mu\mu} = 0$ IH
(Fig.~\ref{fg:C22-IO}) are similar to those for Class $M_{\mu\mu} = 0$
NH (Fig.~\ref{fg:M22-NO}). The similarity of a cofactor-zero IH
scenario with a texture-zero NH scenario can be understood by looking
at the form of the $A$, $B$, and $C$ coefficients in
Table~\ref{tab:ABC}. If we multiply the coefficients for Class
$C_{\mu\mu}=0$ IH by $m_2 m_3$, and divide the coefficients for Class
$M_{\mu\mu}=0$ NH by $m_2 m_3$, we see that $A$ and $B$ become the
same for the two cases. For the $C$ coefficient, the dominant term in
each case is the third one, proportional to $|U_{\mu 3}|^4$ times the
ratio of a larger mass to a smaller one. Therefore the allowed regions
for these two cases are similar. Class $M_{\tau\tau} = 0$ NH and Class
$C_{\tau\tau} = 0$ IH have a similar correspondence in the the $A$,
$B$, and $C$ coefficients, and they are both not allowed for the
best-fit parameters.

Class $M_{ee} = 0$ NH and Class $C_{ee} = 0$ IH also have similar
constraints from the the $A$, $B$, and $C$ coefficients, but there is
no restriction on $\delta$ in those models.  Also, the dominant term
in the $C$ coefficient is suppressed by a factor $|U_{e3}|^4$, so the
other terms become important, and the allowed ranges of the lightest
mass are significantly different in the two models.

One can also see a similarity between cofactor-zero models with NH and
texture-zero models with IH, although the correspondence occurs only
for larger values of the lightest mass. For example, for Class
$C_{\mu\mu}=0$ NH and Class $M_{\mu\mu}=0$ IH, after multiplying the
$A$, $B$, and $C$ coefficients for the NH by $m_2 m_3$ and dividing
the coefficients for the IH by $m_2 m_3$, the $A$ and $B$ coefficients
are the same. When the lightest mass is not too small (such that $m_1
\approx m_2$ for NH), the same terms in the $C$ coefficient are
dominant and proportional to a large mass divided by a small
mass. Thus for higher values of the lightest mass, the allowed regions
of Classes $C_{\mu\mu}=0$ NH and $M_{\mu\mu}=0$ IH must be
similar. This can be seen by comparing Figs.~\ref{fg:C22-NO}
and~\ref{fg:M22-IO}.  However, for small values of the lightest mass,
the first two terms in the $C$ coefficient have similar size for Class
$M_{\mu\mu}=0$ IH, but only the first term is dominant for Class
$C_{\mu\mu}=0$ NH. Thus the allowed regions are quite different when
the lightest mass is below 20~meV.

The allowed regions for Class $M_{\tau\tau} = 0$ IH and Class
$C_{\tau\tau} = 0$ NH also have a similar correspondence in the the
$A$, $B$, and $C$ coefficients, and they have similar allowed regions
for higher values of the lightest mass; see Figs.~\ref{fg:C33-NO}
and~\ref{fg:M33-IO}.  Likewise Class $M_{ee} = 0$ IH and Class $C_{ee}
= 0$ NH have similar $A$, $B$, and $C$ coefficients, and they are both
not allowed at $2\sigma$.

This similarity between texture-zero models with one mass hierarchy and
cofactor-zero models with the other mass hierarchy has been noted
before in models with a single off-diagonal texture or cofactor
zero~\cite{Liao:2013kix}. Thus it is a generic property for any
texture or cofactor zero in the neutrino mass matrix.

\section{Symmetry realization}

All the texture and cofactor zero cases can be realized from discrete
$\mathbb{Z}_N$ symmetries but it requires many scalar
singlets~\cite{Grimus:2004hf}. Here we present a simple realization of
the one cofactor zero models using a new $U(1)$ gauge symmetry that
only requires two scalar singlets. We denote the charge of the new
$U(1)$ gauge symmetry as $Y'$, and make the following charge
assignments: $Y'(q_L)=-Y'(u_R^c)=-Y'(d_R^c)$ for all families in the
quark sector to avoid flavor changing neutral currents;
$Y'(l_{Li})=-Y'(e_{Ri}^c)=-Y'(N_{Ri}^c)$ and $Y'(l_{Li}) \neq
Y'(l_{Lj})$ ($i \ne j$) for each family in the lepton sector; and
$Y'(\phi)=0$ for the SM Higgs. The anomaly-free requirement yields the
condition~\cite{Araki:2012ip}
\begin{equation}
9Y'(q_L)+Y'(l_{L1})+Y'(l_{L2})+Y'(l_{L3})=0.
\end{equation}
If we consider the case with $Y'(q_L)\neq 0$, then the condition leads
to a $B-\sum_\alpha x_\alpha L_\alpha$ gauge symmetry with the
constraint $\sum_\alpha x_\alpha=3$, where $B$ and $L$ are the baryon
and lepton flavor numbers, respectively. One of the advantages of this
model is that both the charged lepton and Dirac neutrino mass matrices
are diagonal spontaneously because of the charge assignments of the
$U(1)$ gauge symmetry. Hence a cofactor zero in $M$ is equivalent to a
cofactor zero in $M_R^{-1}$, which is equivalent to a texture zero in
$M_R$. This can be achieved with a suitable $B-\sum_\alpha x_\alpha
L_\alpha$ gauge symmetry and two SM gauge singlet scalars $S_1$ and
$S_2$ with appropriate charges. Taking the $C_{ee}=0$ case for
example, if we impose a $B-3L_e-L_\mu+L_\tau$ symmetry on the model,
then the $U(1)$ charge matrix for the right-handed neutrino mass term
$Y'(\overline{N}_i^c N_j)$ is:
\begin{align}
Y'=\begin{bmatrix}
   -6 & -4 & -2 \\
   \cdot & -2 & 0  \\
   \cdot & \cdot & 2
   \end{bmatrix}.
\end{align}
Without any additional singlet scalars, the mass matrix of
right-handed neutrinos will only have one non-vanishing entry with the
scale $M_{B-3L_e-L_\mu+L_\tau}$. By adding two additional singlet
scalars $S_1$ and $S_2$ with $|Y'(S_1)|=2$ and $|Y'(S_2)|=4$
respectively, we can make all entries except the $(1,1)$
entry nonzero after $S_1$ and $S_2$ acquire VEVs:
\begin{align}
M_R&=M_{B-3L_e-L_\mu+L_\tau}\begin{bmatrix}
   0 & 0 & 0 \\
   \cdot & 0 & \times  \\
   \cdot & \cdot & 0
   \end{bmatrix}+\left\langle S_1\right\rangle \begin{bmatrix}
      0 & 0 & \times \\
      \cdot & \times & 0  \\
      \cdot & \cdot & \times
      \end{bmatrix}+\left\langle S_2\right\rangle \begin{bmatrix}
            0 & \times & 0 \\
            \cdot & 0 & 0  \\
            \cdot & \cdot & 0
            \end{bmatrix}\nonumber \\
&\sim  \begin{bmatrix}
      0 & \times & \times \\
      \cdot & \times & \times  \\
      \cdot & \cdot & \times
      \end{bmatrix}\,,
      \end{align}
where $\times$ denotes a non-vanishing entry. The other cases can be also realized similarly; a complete list is shown in Table~\ref{tab:realization}.
\begin{table}
\caption{The anomaly-free $U(1)$ gauge symmetry realization for 6 model
classes with one cofactor zero in the light neutrino mass matrix. $Y'$
denotes the charge of the $U(1)$ gauge symmetry, and $S_1$, $S_2$ are
two SM singlet scalars with non-vanishing VEVs.}
\begin{center}
\begin{tabular}{|c|c|c|c|}\hline
Class &Symmetry generator & $|Y'(S_1)|$ & $|Y'(S_2)|$ \\\hline
$C_{ee}$=0&$B-3L_e-L_\mu+L_\tau$&2&4\\\hline
$C_{\mu\mu}$=0&$B+L_e-3L_\mu-L_\tau$&2&4\\\hline
$C_{\tau\tau}$=0&$B-L_e+L_\mu-3L_\tau$&2&4\\\hline
$C_{e\mu}$=0&$B-3L_e-L_\mu+L_\tau$&2&6\\\hline
$C_{\mu\tau}$=0&$B+L_e-3L_\mu-L_\tau$&2&6\\\hline
$C_{e\tau}$=0&$B-L_e+L_\mu-3L_\tau$&2&6\\\hline
\end{tabular}
\end{center}
\label{tab:realization}
\end{table}

\section{Conclusions}

We studied the phenomenology of one diagonal texture or cofactor zero
in the low energy neutrino mass matrix. The cofactor-zero condition is
equivalent to a texture zero in $M^{-1}$ for three massive
neutrinos. In the case that the Yukawa coupling matrix is diagonal, a
texture zero in $M$ is equivalent to a cofactor zero in $M_R$, and a
cofactor zero in $M$ is equivalent to a texture zero in $M_R$. We
imposed one diagonal texture or cofactor zero on the neutrino mass
matrix and used the latest experimental data to obtain the allowed
regions for the lightest neutrino mass and Dirac CP phase $\delta$.
The texture zero cases $M_{\tau\tau}=0$ NH and $M_{ee}=0$ IH, and the
cofactor zero case $C_{ee}=0$ NH are not allowed at $2\sigma$, and
the case $C_{\tau\tau}=0$ IH is not allowed for the best-fit parameters.

Once the lightest neutrino mass and Dirac CP phase were determined, we
made definite predictions for neutrinoless double beta decay for
one texture or cofactor zero models. The effective mass $|M_{ee}|$ is
generally proportional to the lightest mass (see the iso-$|M_{ee}|$
contours in Figs. 1-6), which is clearly evident for the
quasi-degenerate spectrum. However, $|M_{ee}|$ is strongly dependent
on the Dirac CP phase $\delta$ when the lightest mass
is small, of order 20 meV or less.


%
\begin{table}
\caption{The minimum values of $|M_{ee}|$ (in $10^{-3}$ eV) in each
  class for the best-fit oscillation parameters, and the $2\sigma$
  lower bounds. The symbol $\times$ denotes that there is no allowed
  region for the model.}
\begin{center}
\begin{tabular}{|c|c|c|c|c|}\hline
\multirow{2}{*}{Class}&\multicolumn{2}{|c|}{Best-fit}&\multicolumn{2}{|c|}{$2\sigma$ lower bound}\\\cline{2-5}
                      &NH&IH&NH&IH\\\hline
$M_{ee}$=0&0.0&$\times$&0.0&$\times$\\\hline
$M_{\mu\mu}$=0&34.4&19.1&26.8&15.1\\\hline
$M_{\tau\tau}$=0&$\times$&18.2&$\times$&14.8\\\hline
$C_{ee}$=0&$\times$&18.1&$\times$&14.8\\\hline
$C_{\mu\mu}$=0&0.0&39.7&0.0&29.6\\\hline
$C_{\tau\tau}$=0&0.0&$\times$&0.0&32.3\\\hline

\end{tabular}
\end{center}
\label{tab:Mee}
\end{table}

The minimum value of $|M_{ee}|$ for the best-fit oscillation
parameters and the $2\sigma$ lower bounds for the diagonal cases are
shown in Table~\ref{tab:Mee}. Results for the off-diagonal cases can
be found in Ref.~\cite{Liao:2013kix}. For the diagonal cases that are
not excluded, the minimum $|M_{ee}|$'s are all below $40$~meV and can
only be completely probed by significant improvements in the
sensitivity of $0\nu\beta\beta$ experiments. For Classes $M_{ee}=0$
NH, $C_{\mu\mu}=0$ NH and $C_{\tau\tau}=0$ NH, the current lower bound on
$|M_{ee}|$ is zero. 

However, as we have shown, for larger values of the lightest neutrino
mass (especially in the quasi-degenerate region) the similarity of the
allowed regions between a texture-zero NH and the corresponding
cofactor zero-IH (and a texture-zero IH and the corresponding
cofactor-zero NH) makes it difficult to distinguish them simply with
measurements of the oscillation parameters and the neutrino mass
scale. In order to resolve this ambiguity, future experiments that can
determine the mass hierarchy are strongly needed, such as long
baseline neutrino experiments (T2K~\cite{Abe:2011ks},
NO$\nu$A~\cite{Patterson:2012zs}, and LBNE~\cite{Akiri:2011dv}), atmospheric
neutrino experiments (PINGU~\cite{Clark:2012hya} and
INO~\cite{Mondal:2012xm}) and medium baseline reactor experiments (Daya Bay
II~\cite{dayabayii:dayabayii}).

\section*{Acknowledgments}

JL and KW thank the University of Kansas for its hospitality during the
initial stages of this work. DM thanks the Kavli Institute for
Theoretical Physics for its hospitality during the completion of this
work. This research was supported by the U.S. Department of Energy
under Grant Nos. DE-FG02-01ER41155, DE-FG02-04ER41308, and
DE-FG02-13ER42024, and by the National Science Foundation under
grant No. PHY11-25915.

\newpage

\end{document}